\documentclass{emulateapj}   
\usepackage{apjfonts}

\submitted{ApJL, accepted (30 May 2007)}

\def\ltsima{$\; \buildrel < \over \sim \;$}
\def\simlt{\lower.5ex\hbox{\ltsima}} 
\def\gtsima{$\; \buildrel > \over \sim \;$}
\def\simgt{\lower.5ex\hbox{\gtsima}} 
\def\arcsec{\hbox{$^{\prime\prime}$}}
\def\deg{\hbox{$^\circ$}}
\def\HST{{\it HST}}
\def\VLA{{\it VLA}}
\def\VLBA{{\it VLBA}}
\def\Chandra{{\it Chandra}}

\shorttitle{Superluminal Motion in HST-1} 
\shortauthors{Cheung, Harris, \& Stawarz}

\begin{document}

\title{Superluminal Radio Features in the M87 Jet and the Site of Flaring
TeV Gamma-ray Emission}

\author{C.~C. Cheung\altaffilmark{1,2}, D.~E. Harris\altaffilmark{3}, \L. 
Stawarz\altaffilmark{2,4}}

\altaffiltext{1}{Jansky Postdoctoral Fellow of the National Radio
Astronomy Observatory; teddy3c@stanford.edu.}

\altaffiltext{2}{Kavli Institute for Particle Astrophysics and Cosmology, 
Stanford University, Stanford, CA 94305; stawarz@slac.stanford.edu.}

\altaffiltext{3}{Harvard-Smithsonian Center for Astrophysics, 60 Garden St.,
Cambridge, MA 02138; harris@cfa.harvard.edu.}

\altaffiltext{4}{Astronomical Observatory, Jagiellonian University, ul. 
Orla 171, 30-244 Krak\'ow, Poland.}

\begin{abstract} 

Superluminal motion is a common feature of radio jets in powerful
$\gamma$-ray emitting active galactic nuclei. Conventionally, the variable
emission is assumed to originate near the central supermassive black-hole
where the jet is launched on parsec scales or smaller.  Here, we report
the discovery of superluminal radio features within a distinct flaring
X-ray emitting region in the jet of the nearby radio galaxy M87 with the
{\it Very Long Baseline Array}. This shows that these two phenomenological
hallmarks -- superluminal motion and high-energy variability -- are
associated, and we place this activity much further ($\geq$120 pc) from
the ``central engine'' in M87 than previously thought in relativistic jet
sources. We argue that the recent excess very high-energy TeV emission
from M87 reported by the H.E.S.S. experiment originates from this variable
superluminal structure, thus providing crucial insight into the production
region of $\gamma$-ray emission in more distant blazars. 

\end{abstract}


\section{Introduction\label{section-intro}}

\begin{figure*}
\epsscale{1.05}
\plotone{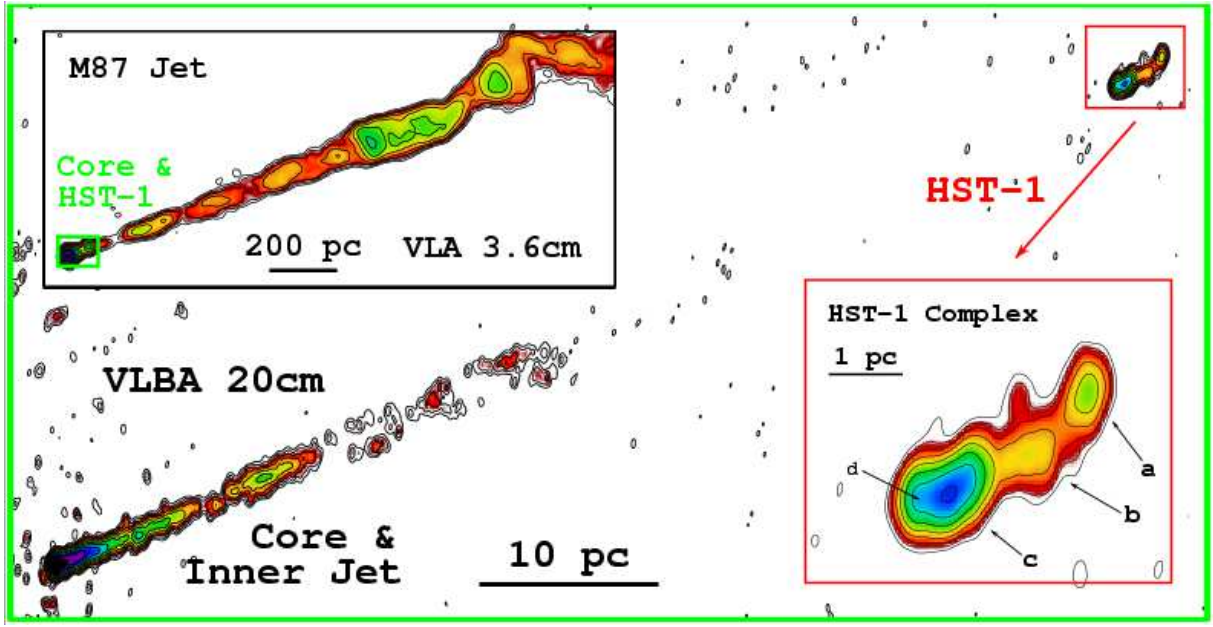}
\caption{
Multi-scale images of the M87 jet.
Our \VLBA\ images cover the inner $\sim$1\arcsec\ of the jet, as
indicated by the corresponding green box in our VLA 3.6cm image (from Dec. 
2004; upper left inset) of the large-scale jet. The bottom right inset is a zoom-in of the HST-1
complex as imaged with the \VLBA\ (1 pc = 13
mas); the lowest contour is 0.3 mJy/bm.}
\label{fig-1}
\end{figure*}

The proximity of M87 \citep[$D$=16 Mpc;][]{ton91} makes it one of the best
systems to study relativistic jets at high linear resolution
(Fig.~\ref{fig-1}).  Our observations with the \Chandra\ X-ray Observatory
isolated short (month) timescale variability \citep{har06} in a jet region
previously dubbed `HST-1', which is separated by 0.86\arcsec\ (60 pc,
projected) from the central ``active'' supermassive black hole (SMBH). 
The variability culminated in a factor $>$50 X-ray outburst making HST-1
the brightest X-ray source in the galaxy for a few years. 
Figure~\ref{fig-2} shows a lightcurve from 2000 to the end of 2006. 

Observations with the {\it Hubble Space Telescope} (\HST) and the {\it
Very Large Array} (\VLA) show comparable activity in the optical and at
radio frequencies \citep{per03,har06}.  After the detection of appreciable
radio flux in HST-1 from our first season of \VLA\ observations (in 2003),
we began to monitor the jet at higher (sub-parsec) resolution with the
NRAO\footnote{The National Radio Astronomy Observatory is operated by
Associated Universities, Inc. under a cooperative agreement with the
National Science Foundation.} {\it Very Long Baseline Array} (\VLBA) at
three frequencies (0.33, 0.61, and 1.7 GHz) commencing in Jan. 2005.  This
was the period at which the X-ray and optical intensities were peaking,
and the radio intensity plateaued (Fig.~\ref{fig-2}).  Here, we report on
the highest resolution (1.7 GHz) observations which resolve dynamic
structures within the flaring X-ray region, including the discovery of
superluminal motion in multiple knots. The implications for observations
of more distant $\gamma$-ray emitting relativistic jets are discussed.

\section{Description of Observations}

\begin{figure}
\epsscale{1.1}
\plotone{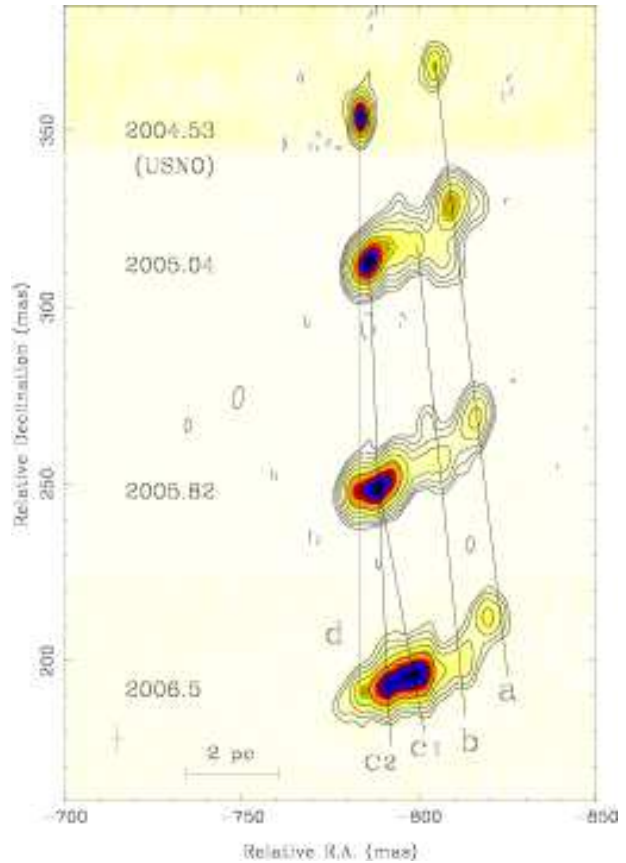}
\caption{Representative sequence of \VLBA\ images
of knot HST-1 (see Fig.~\ref{fig-1}) restored with
a 8.0 mas $\times$ 3.4 mas (0.62 pc $\times$
0.26 pc) beam at PA=3\deg\ (bottom left). The
vertical spacing is proportional to time elapsed and
straight lines trace the motions of the different features (note the
apparent deceleration of knot ``a'').  The first epoch is a shorter 2.3   
GHz dataset resulting in the excess background; subsequent
ones are full track 1.7 GHz observations. }
\label{fig-3}
\end{figure}

Each of our nine \VLBA\ runs consisted of six one-hour integrations of M87
over an 8 hr period, with calibrator observations interleaved.  Four
adjacent 8 MHz bandwidth channels were used, centered at 1.667 GHz.  The
full 10 antenna array was used except in Jan. 2005 (missing Hancock
antenna), May 2005 (missing Saint Croix), and May 2006 (missing Pie Town).
The data were calibrated using NRAO's \VLBA\ data calibration pipeline in
AIPS and post-processed with a combination of DIFMAP and AIPS. The rms
noise in the \VLBA\ images are within 15$\%$ of the theoretically
predicted value for the full \VLBA\ (0.053 mJy/bm). The images have been
restored with a common beam (Fig.~\ref{fig-3}) which is within 10$\%$ of
the uniformly weighted beams of all but the May 2005 epoch. 

Over the inner jet, the new sequence of \VLBA\ images shows previously
known radio structure out to $\sim$250 mas from the core
\citep{bir95,dod06} with additional transversely resolved emission between
$\sim$200-450 mas (Fig.~\ref{fig-1}). The central position angle
(PA$\sim$290\deg) of the first $\sim$100 mas portion of the jet aligns
roughly with the large-scale \VLA\ jet.  Knot `HST-1' is further
downstream and is offset from the central jet being aligned with the
projection of the northern edge of the resolved $\sim$200-450 mas jet
(i.e., PA$\approx$294.5\deg). 

A short 2.3 GHz \VLBA\ observation obtained in Jul. 2004 by the
USNO\footnote{This research has made use of the United States Naval
Observatory Radio Reference Frame Image Database (RRFID)} \citep{fey00}
detected only the two brightest features in HST-1 seen in our early \VLBA\
images allowing us to trace these structures back by $\sim$1/2 year before
our program commenced. Further archival VLBI observations of varying
quality as far back as in 2000 \citep[e.g.,][]{dod06} did not give any
significant detections of HST-1. We registered the images from the
different epochs on the position of the maximum of the radio core
(Fig.~\ref{fig-1}).  Based on the stability of the structure of the inner
jet \citep{dod06} over the 10 total epochs of observations, the core
position is determined to be aligned to a fraction of the common beam over
the two year period. 

With the images registered, the upstream (eastern) edge which we call
`HST-1d', is the dominant feature in the HST-1 complex at the early
epochs.  The later epoch images reveal the emergence of a radio knot
(HST-1c) moving downstream from HST-1d at a rate of 4.48$\pm$0.42 mas/yr
($\beta_{\rm app}=1.14 \pm 0.14 c$) at a PA of 279\deg\ and the peak radio
surface brightness decays by only $\sim$20$\%$ over a 1 year period.
Sometime between Dec. 2005 and Feb. 2006, knot HST1c evidently splits into
two roughly equally bright features: a faster moving component (c1;
4.3$\pm$0.7$c$) and a slower moving trailing feature (c2;
0.47$\pm$0.39$c$). Between the two years, the location of HST-1d is
basically stationary to within $\sim$2 mas (i.e., its motion is
$<$0.25$c$) at 860 mas from the core and is the apparent point of origin
of the superluminal ejections. 

\begin{figure}
\epsscale{1.15}
\plotone{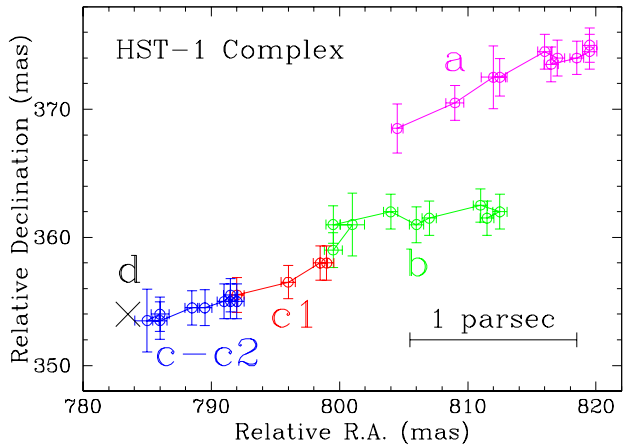}
\caption{Positions of radio features in the
HST-1 complex relative to the core (at origin) from our \VLBA\ data (see
Fig.~\ref{fig-3}). The positional errors are
1/6th of the uniform weighted beam-sizes. Component `c' emanates from the 
basically stationary feature
`d' (black cross) and splits into c1 and c2.
All components are moving downstream (upper right).}
\label{fig-4}
\end{figure}

The most distant knot (HST-1a) is well-isolated from the other structures
in every observation. We observe it moving downstream at 2.49$\pm$0.25$c$
at PA = 295\deg$\pm$8\deg\ (basically radial) until Dec. 2005 when it
appears to decelerate to 1.41$\pm$0.49$c$ at a smaller PA of 289\deg\
(although the PA change is apparent in Figure~\ref{fig-4}, it is not
statistically significant). A fainter feature (HST-1b), identifiable
beginning 2005, trails HST-1a at an identical speed (2.52$\pm$0.14$c$)
with a non-radial trajectory (a smaller PA of 279\deg$\pm$6\deg).  Feature
c1 (see above) actually ends up in Jul. 2006 where HST-1b was first
detected in Jan. 2005. 

Previous reports of motions up to $\sim$6c from HST-1 were based on yearly
\HST\ optical monitoring \citep{bir99} from 1994 to 1998 but pertained to
emission further downstream from the \VLBA\ structures discussed here.
Furthermore, our motions are apparent on month timescales rather than
years because of the higher spatial resolution of our \VLBA\ data. The
extent of the radio emission in HST-1 is only $\sim$40 mas, roughly 3
pixels across in the \citet{bir99} \HST\ FOC images; a \Chandra\ ACIS
pixel is $\sim$30$\times$ larger.

\section{Discussion and Summary}

The importance of the superluminal motions reported here is twofold. First
and foremost, the HST-1 complex is well-isolated from the nucleus and the
rest of the jet, so the outbursting higher-energy (X-ray, optical)
emission isolated by \Chandra\ and \HST\ can be uniquely attributed to the
region resolved by our \VLBA\ observations (Fig.~\ref{fig-1}).  While VLBI
detections of superluminal motions are now commonplace in more distant
relativistic jet sources \citep[e.g.,][]{kel04}, the physical link to the
higher-energy activity has not been possible previously because of the
lack of comparably high spatial resolution \citep[e.g.,][]{jor01}. Second,
contrary to conventional wisdom, this `blazar'-like activity is clearly
displaced from the central engine by $\geq$120 pc (deprojected; see
\S~\ref{section-3.2}). Thus, at least in the case of M87, the observed
hallmarks of blazar behavior are not directly associated with the
immediate vicinity of the SMBH where the jet is launched \citep{jun99}. 
Without the comparably high linear resolution, an analogous sequence of
events in a more distant source would be associated with the base (i.e.,
sub-pc to few-pc scales) of the jet where the `non-jetted' contribution
from the active galactic nucleus (AGN) to the observed X-rays \citep[such
as from the accretion disk;][]{mar02} would contaminate the lightcurves.

\subsection{The Origin of VHE $\gamma$-ray Emission in M87}

\begin{figure}
\epsscale{1.15}
\plotone{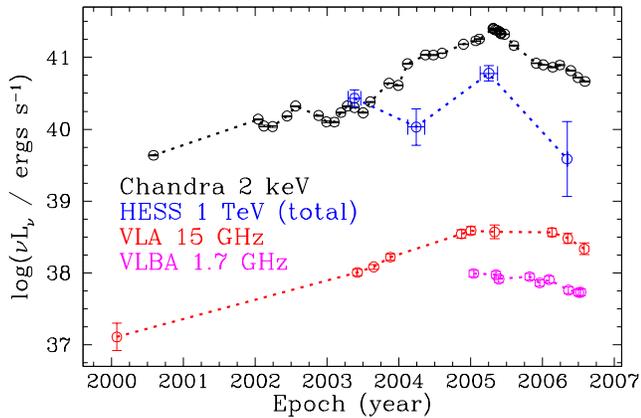}
\caption{Lightcurves of the total TeV
emission from M87 \citep[taken from][]{aha06} and of the jet knot
HST-1 (X-ray and two radio bands).  The 2 keV and 15 GHz data
up to $\sim$Aug. 2005 were previously published in \citet{har06}. Our  
subsequent observations now show the X-ray intensity of HST-1
declining steeply, similar to the total TeV emission from M87. The 1.7 GHz
\VLBA\ points are integrated from the entire HST-1 complex.} \label{fig-2}
\end{figure}

At even higher-energies, the H.E.S.S. collaboration recently reported
flaring TeV emission from M87 \citep{aha06}.  This emission revealed
gradual (year timescale) variability, with a maximum coinciding with the
peak of the radio-to-X-ray activity detailed in HST-1 (mid-2005;
Fig.~\ref{fig-2}) suggesting a link between these flares.  However, the
HST-1 knot was dismissed as a possible production site of the TeV emission
\citep{aha06} because of the short (days) timescale TeV variability
detected on top of the longer timescale variability. Such rapid variations
were considered unlikely for HST-1 since they imply (through the causality
argument) an emission region size\footnote{The Doppler factor is $\delta =
1 / \Gamma \, (1 - \sqrt{1-\Gamma^{-2}} \, \cos \theta)$, where $\Gamma$
is the bulk Lorentz factor, and $\theta$ is the angle to the line of
sight.} $R_{\rm var} \simlt 0.002 \, \delta$\,pc.  We can neither claim
nor reject the presence of $\sim$day time-scale variability in our optical
or X-ray data for HST-1 due to insufficient sampling. However, the \VLBA\
data show the compact knots in HST-1 to be essentially unresolved with
semi-minor axes $<$0.15 pc, and the current optical/X-ray variability data
constrain $R_{\rm var} \simlt 0.022 \, \delta$\,pc (see below),
approaching the size limits set by the variability of the TeV emission. 

Here, we suggest HST-1 as a plausible site for the production of a
dominant portion of the detected VHE TeV emission due to inevitable
inverse-Compton upscattering off ambient photon fields (e.g., starlight)
by the electrons producing the flaring synchrotron optical-to-X-ray
emission. This was in fact proposed by \citet{sta06} in discussing earlier
(2004 and before) TeV detections of M87.  The connection is strengthened
by the fact that the maxima of the total TeV flux density from M87 and
synchrotron (HST-1 only) flares observed in 2005 were coincident, and that
their luminosities are comparable (Fig.~\ref{fig-2}). In addition, the
$0.4-10$\,TeV emission from M87 at this time is well described by a single
power-law with spectral index $\alpha_{\gamma} = 1.2\pm0.15$
($S_{\nu}\propto\nu^{-\alpha}$), while the optical-to-X-ray power-law
slope of HST-1 during the same period is similar: $\alpha_{\rm OX} =
0.99\pm0.03$.  These observations are consistent with a common origin for
the flaring radio through $\gamma$-ray emission from HST-1 as outlined in
more detail in \citet{sta06}.

The above conclusion is supported by the lack of any other plausible
production site for the VHE $\gamma$-ray emission in M87.  For example,
the innermost (sub-pc scale) jet region is characterized by only
small-amplitude optical/X-ray variability \citep{har06} and only mildly
relativistic radio features \citep{ly07}. As argued also by \citet{aha06},
this is inconsistent with models for the generation of VHE emission in the
unresolved core \citep{geo05}. Instead, \citet{aha06} considered curvature
radiation of ultra-high energy protons \citep{lev00,bol00} accelerated by
a strong magnetic field in the closest vicinity ($\sim 3 \, R_g$) of the
SMBH at the center of M87.  However, this interpretation is problematic
also due to the fact that the nearest environments of active SMBHs are
expected to be opaque to TeV photons due to photon-photon ($\gamma\gamma$)
annihilation on ambient photon fields such as from the accretion disk. 


In M87, the $\gamma$-ray photons in the energy range covered by H.E.S.S.
($\varepsilon = 1-10$ TeV) interact mostly with photons emitting at $\nu_0
\sim 2 m_e^2 c^4/h \varepsilon \sim 10^{13-14}$\,Hz, i.e., infrared to
optical ones. Quantitatively, the optical depth for the $\gamma\gamma$
annihilation process can be approximated by $\tau_{\gamma\gamma} \approx
(1/3) \sigma_{\rm T} \, r \, n_0$, where $\sigma_{\rm T}$ is the Thomson
cross-section.  The number density $n_0$ of the $h \nu_0$ nuclear photons
within a spherical volume around the SMBH of radius $r$ is related to the
monochromatic luminosity $L_{\rm 0}$ by $h \nu_0 \, n_0 = L_{\rm 0} / 4
\pi \, r^2 c$ (the main contributor to the photon energy density is judged
to be from the accretion disk). Hence one can find, $\tau_{\gamma\gamma}
\approx 200 \, \left(L_{\rm 0} / 10^{40} \, {\rm erg \, s^{-1}}\right) \,
\left({\nu_0 / 10^{13} \, {\rm Hz}}\right)^{-1} \, \left({r /
R_g}\right)^{-1}$.  The considered region would be transparent to the TeV
photons only if $\tau_{\gamma\gamma} < 1$.  However, at $r \sim (3-10)
\times R_g$ we expect $\tau_{\gamma\gamma} \gg 1$, since the observed
luminosity of the M87 nucleus \citep{kha04,why04} at $\nu_0 =
10^{13-14}$\,Hz is $L_{\rm 0} \simgt 3 \times 10^{40}$\,erg\,s$^{-1}$,
consistent with ADAF approximations of the accretion disk \citep{dim03}. 
With the central region of the AGN excluded as a plausible site of the TeV
emission, the flaring HST-1 knot is the most probable point of origin.
Therefore, in addition to the flaring synchrotron (radio-to-X-ray)
emission, the superluminal radio structures detailed here can also be
associated with VHE $\gamma$-ray activity.

\subsection{HST-1 as a Recollimation Shock\label{section-3.2}}

A natural question to ask is if the position of the HST-1 knot in the M87
jet is `special' in some way. In other words, what determines the location
and flaring behavior of this extremely compact and variable feature within
the relativistic outflow?  The kinematics of the superluminal radio
components (in particular, the 4.3c motion in c1) constrain the jet at the
location of HST-1 to be aligned at $\theta \leq 26 \pm 4 \deg$, and
requires $\Gamma \geq \sqrt{\beta_{\rm app}^{2} + 1} = 4.4$, and $\delta
\simgt 2$.  The HST-1 region is therefore $\geq$120 pc (deprojected
assuming $\theta=30\deg$). For a SMBH with $\mathcal{M}_{\rm BH} \approx 3
\times 10^9 \, M_{\odot}$ \citep{mac97}, this corresponds to the distance
of $\simgt 10^6$ Schwarzchild radii ($R_g = G \mathcal{M}_{\rm BH} / c^2
\approx 4.4 \times 10^{14}$\,cm $\approx 1.4 \times 10^{-4}$\,pc). At such
a large distance from the active center, our X-ray observations establish
flux doubling timescales $t_{\rm var}$ of 0.14 yrs \citep{har06}, which
constrains $R_{\rm var} < c \, t_{\rm var} \, \delta \approx 0.022 \,
\delta$\,pc. 

Recently, \citet{sta06} showed that the position of the HST-1 knot agrees
with the expected location of a `re-confinement nozzle' formed within the
M87 jet due to a converging shock driven by the interaction of the outflow
with the interstellar medium (ISM). In this ambient medium, it was
postulated that the gravitational influence of the central SMBH in the
inner region of the M87 host galaxy \citep{lau92} results in an increase
of the gas pressure, analogous to the forming of the observed enhanced
stellar cusp \citep{you78}.  For the temperature $kT_{\rm ism} \approx
0.8$\,keV of hot gaseous ISM (with the number density $n_{\rm ism} \approx
0.17$\,cm$^{-3}$) in the inner ($<1$\,kpc) parts of the galaxy, the
gravitational capture radius is $R_A = G \mathcal{M}_{\rm BH} / c_s^2
\approx 100$\,pc, where $c_s$ is the appropriate sound speed
\citep{dim03}.  It turns out that $\sim 100$\,pc is the spatial scale of
the disk of ionized gas observed in M87 by the \HST\ \citep{for94}. It was
the Keplerian rotation of the inner parts of this disk which enabled a
precise estimate of the black hole mass in this system \citep{mac97}. 

Thus, one can surmise that a gravitationally perturbed ambient medium
leads naturally to the formation of a jet feature like HST-1 at $\simgt
100$\,pc from a central SMBH. In this scenario, the `stationary' region
HST-1d defines the opening of the nozzle. The origins of the different
superluminal features can be traced back to this position at $\sim$2004.5
(HST-1c), 2003.3 (HST-1b), and 2001.9 (HST-1a). The latter two could have
originated more recently as they assume the nominal velocities of 2.5$c$;
the apparent curve in trajectory and deceleration of HST-1a lead us to
believe it could have been moving faster previous to our first \VLBA\
imaging observations. 

The formation of a `reconfinement nozzle' within the jet is not unique for
the hydrodynamical outflow considered by \citet{sta06}.  It was previously
shown \citep{tsi02} that interactions between a strongly magnetized
relativistic outflow with a non-relativistic collimating
magneto-hydrodynamical (MHD) wind may lead as well to the formation of a
converging shock within the relativistic component.  An evaluation of the
exact position of this shock from the jet base is extremely
model-dependent. Nevertheless, for the jet parameters anticipated by
\citet{gra05}, who successfully explained the observed gradual collimation
of the M87 jet between $10^{2} - 10^{6} \, R_g$ \citep{jun99} based on the
\citet{tsi02} MHD model, the expected position is near the location of the
HST-1 knot. We speculate that a rapid release of the magnetic energy
within the formed nozzle via magnetic reconnection may play an important
role in producing the observed high energy flares and ejection of the
relativistic radio blobs, somewhat analogous to coronal mass ejections
from the Sun \citep{pic06}. 


In summary, we have discovered superluminal motions of radio features in
the M87 jet at a site remote from the central SMBH.  These features appear
to be associated with the remarkable flare of 2005 for which the radio,
optical, and X-ray flux densities peaked at levels of $>$30--50 times that
of a few years earlier.  We have argued that the TeV excess intensity of
2005 detected from M87 by H.E.S.S. was a manifestation of the same event. 
Thus, most of the defining characteristics of blazars have now been shown
to occur at a distance $\geq$120 pc from the SMBH instead of from the
immediate environs of the central engine, as is commonly believed to be
the case.  As we enter the {\it GLAST} era, ensuing studies of high-energy
flares from blazars should consider such a production site as we have
resolved in M87. 

\acknowledgments

We thank Bill Junor for his involvement in the early stages of this
project, Roger Blandford for discussions on the $\gamma$-ray opacity, and
Al Marscher and Ken Kellermann (the referee) for helpful comments on the
manuscript. This work was supported by NASA (D.~E.~H.) and by MEiN through
research project 1-P03D-003-29 from 2005-2008 (\L.~S.).

\end{document}